\documentclass[journal, onecolumn, 12pt]{IEEEtran}
\renewcommand{\baselinestretch}{2.0}
\usepackage{cite}
\usepackage{ifpdf}
\usepackage{array}
\ifpdf
\usepackage{graphicx}
\usepackage[svgnames]{xcolor}
\else
\usepackage[dvipdfmx]{graphicx}
\usepackage[dvipdfmx,svgnames]{xcolor}
\fi
\usepackage{amsmath,amssymb,amscd}
\usepackage{verbatim}

\newtheorem{rem}{Remark}

\usepackage{algorithm}
\usepackage{algpseudocode}

\begin{document}
\renewcommand{\baselinestretch}{1.55}
\title{Fronthaul Compression and Precoding Design \\ for C-RANs over Ergodic Fading Channels}

\author{\large Jinkyu Kang, Osvaldo Simeone, Joonhyuk Kang and Shlomo Shamai (Shitz)
\thanks{Jinkyu Kang and Joonhyuk Kang are with the Department of Electrical Engineering, Korea Advanced Institute of Science and Technology (KAIST) Daejeon, South Korea (Email: kangjk@kaist.ac.kr and jhkang@ee.kaist.ac.kr).

O. Simeone is with the Center for Wireless Communications and Signal Processing Research (CWCSPR), ECE Department, New Jersey Institute of Technology (NJIT), Newark, NJ 07102, USA (Email: osvaldo.simeone@njit.edu). 

S. Shamai (Shitz) is with the Department of Electrical Engineering, Technion, Haifa, 32000, Israel (Email: sshlomo@ee.technion.ac.il).
}}
\maketitle
\renewcommand{\baselinestretch}{2.0}
\begin{abstract}
This work investigates the joint design of fronthaul compression and precoding for the downlink of Cloud Radio Access Networks (C-RANs). In a C-RAN, a central unit (CU) performs the baseband processing for a cluster of radio units (RUs) that receive compressed baseband samples from the CU through low-latency fronthaul links. Most previous works on the design of fronthaul compression and precoding assume constant channels and instantaneous channel state information (CSI) at the CU. This work, in contrast, concentrates on a more practical scenario with block-ergodic channels and considers either instantaneous or stochastic CSI at the CU. Moreover, the analysis encompasses both the  {\textit{Compression-After-Precoding}} (CAP) and the {\textit{Compression-Before-Precoding}} (CBP) schemes. With the CAP approach, which is the standard C-RAN solution, the CU performs channel coding and precoding and then the CU compresses and forwards the resulting baseband signals on the fronthaul links to the RUs. With the CBP scheme, instead, the CU does not perform precoding but rather forwards separately the information messages of a subset of mobile stations (MSs) along with the compressed precoding matrices to the each RU, which then performs precoding. Optimization algorithms over fronthaul compression and precoding for both CAP and CBP are proposed that are based on a stochastic successive upper-bound minimization approach. Via numerical results, the relative merits of the two strategies under either instantaneous or stochastic CSI are evaluated as a function of system parameters such as fronthaul capacity and channel coherence time. 
\end{abstract}

\newpage
\begin{IEEEkeywords}
Cloud-Radio Access Networks (C-RAN), MIMO, fronthaul compression, stochastic CSI, precoding.
\end{IEEEkeywords}

\section{Introduction}
As industry and academia reconsider conventional cellular systems in the face of unprecedented wireless traffic growth, the Cloud-Radio Access Network (C-RAN) architecture has emerged as a promising solution due to its potential to overcome the problems of cell association and interference management \cite{ChinaMobile, FlanaganTI2011, Marsch12VTMAG, LightradioAL}. In a C-RAN, a dense deployment of radio units (RUs) is made possible by the centralized control performed at central units (CUs), which are connected to a cluster of RUs via low-latency fronthaul links. This control encompasses all protocol layers including the baseband signal level at the physical layer. However, the large bit rate requirement of the digitized baseband signals that are exchanged on the fronthaul links, poses a serious limitation to the feasibility of C-RANs and has motivated significant work on the design of fronthaul compression strategies \cite{Samardzija12TWC, Nieman13GlobalSIP}.

Focusing on the downlink, the standard C-RAN solution prescribes all baseband processing to be performed at the CU on behalf of all connected RUs. Accordingly, the CU compresses the processed baseband signals and forwards them on the fronthaul links to the corresponding RUs. Then, the RUs upconvert and transmit the compressed baseband signals to the mobile stations (MSs). This approach, which is referred to here as a {\textit{Compression-After-Precoding}} (CAP), is studied in, e.g., \cite{Simeone09EURADVSP, Marsch09GLOBECOM, Yu14ITA, Park14SPMAG, Park13TSP}. According to an alternative strategy known as a  {\textit{Compression-Before-Precoding}} (CBP) \cite{Chae13ICC}, the CU still calculates the precoding matrices, but it does not encode and precode the data streams; rather, it forwards the data streams and the precoding matrices to the RUs, which then perform encoding and precoding. A hybrid technique between CAP and CBP is also potentially advantageous as suggested by \cite{Yu14ITA}. 

In previous works \cite{Simeone09EURADVSP, Marsch09GLOBECOM, Yu14ITA, Park14SPMAG, Park13TSP, Chae13ICC}, the design of fronthaul compression and precoding was mostly dealt under the assumption of static channels and full channel state information (CSI) at the CU \cite{Sanderovich09TIT}. This work is instead motivated by the increasing relevance, in modern cellular systems, of channel models that encompass multiple channel coherence blocks within each coding block \cite{LozanoCOMMAG2012}. An example is given by the LTE standard in which a codeword spans multiple resource blocks in the time-frequency domain  \cite{AndrewsBook}. Furthermore, in such systems, full CSI is practically difficult to achieve due to the channel variability within the coding block. For these reasons, we adopt a block-ergodic fading model, in which each codeword spans multiple finite-duration channel coherence blocks, as in, e.g., \cite{Hassibi03TIT, Kobayashi11TSP}. Moreover, we consider both the ideal case of perfect instantaneous CSI and a set-up in which the CU only has stochastic CSI, namely information about the spatial correlation of the channels, as in, e.g., \cite{Shi14ICC, SSUM_paper}. We investigate the joint design of fronthaul compression and precoding for both CAP and CBP strategies. To this end, we leverage information-theoretic bounds on the compression rates (see \cite{Sanderovich09TIT, Park14SPMAG, Park13TSP, Kang14TWC, Zhang14arXiv, GamalBook}) and tackle the optimization problem of maximizing the ergodic capacity for both CAP and CBP. With stochastic CSI, we propose an algorithm based on the Stochastic Successive Upper-bound Minimization (SSUM) scheme \cite{SSUM_paper} that is known to have guaranteed convergence to a local optimum. We provide a thorough performance comparison between the CAP and CBP schemes via numerical results, illustrating the relative merits of the two techniques as a function of system parameters such as fronthaul capacity and channel coherence time, and discuss the impact of stochastic CSI as compared to full CSI.

The rest of the paper is organized as follows. We describe the system model in Section \ref{Sec:SM}. In Section \ref{Sec:PBC}, we study the CAP strategy, while the CBP approach is studied in \ref{Sec:PBC_CBP}, respectively. In Section \ref{Sec:Numerical Results}, numerical results are presented. Concluding remarks are summarized in Section \ref{Sec:Conclusion}.

\emph{Notation}: $E[ \cdot ]$, $\textrm{tr}( \cdot )$, and $[ \cdot]_{i,j}$ denote the expectation, trace and element $(i, j)$ of the argument matrix, respectively. We use the standard notation for mutual information \cite{GamalBook}. ${\bf{\nu}}_{\textrm{max}}^{(d)} ({\bf{A}})$ is a unitary matrix containing as columns the $d$ eigenvectors to the largest eigenvalues of the semi-positive definite matrix ${\bf{A}}$. We reserve the superscript ${\bf{A}}^{T}$ for the transpose of ${\bf{A}}$,  ${\bf{A}}^{\dagger}$ for the conjugate transpose of ${\bf{A}}$, and ${\bf{A}}^{-1} = ({\bf{A}}^\dagger {\bf{A}})^{-1} {\bf{A}}^\dagger$, which reduces to the usual inverse if the number of columns and rows are same. The $n \times n$ identity matrix is denoted as ${\bf{I}}_n$.
\section{System Model} \label{Sec:SM}
\begin{figure}[t]
\centering
\includegraphics[width=15cm]{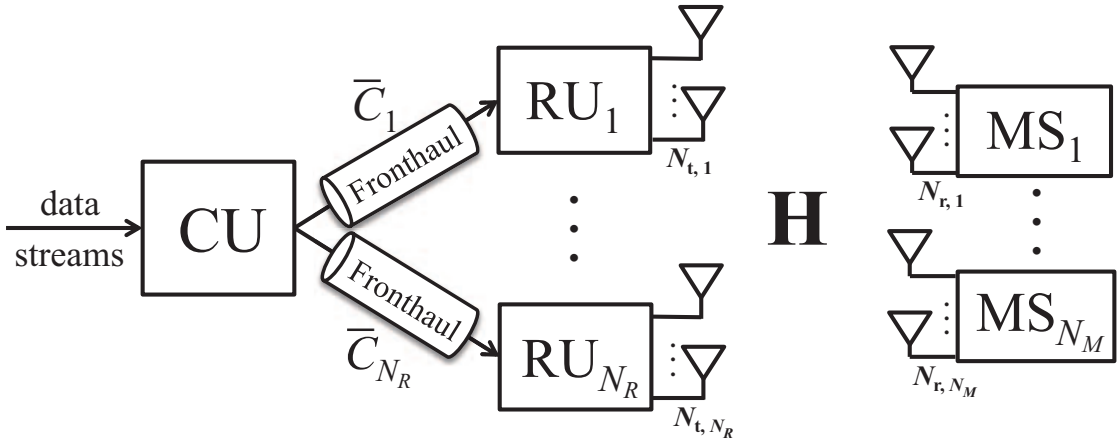}
\caption{Downlink of a C-RAN system in which a single cluster of RUs is connected to a CU via finite-capacity fronthaul links. The downlink channel matrix ${\bf{H}}$ varies in an ergodic fashion along the channel coherence blocks.}
\label{fig:fig1}
\end{figure}

We consider the downlink of a C-RAN in which a cluster of $N_R$ RUs provides wireless service to $N_M$ MSs as illustrated in Fig. \ref{fig:fig1}. Most of the baseband processing for all the RUs in the cluster is carried out at a CU that is connected to each $i$-th RU via a fronthaul link of finite capacity, as further discussed below. Each $i$-th RU has $N_{t,i}$ transmit antennas and each $j$-th MS has $N_{r,j}$ receive antennas. We denote the set of all RUs as $\mathcal{N}_R = \{1, \dots, N_R\}$ and of all MSs as $\mathcal{N}_M = \{1, \dots, N_M\}$. We define the number of total transmit antennas as $N_t = \sum_{i=1}^{N_R} N_{t,i}$ and of total receive antennas as $N_r = \sum_{j=1}^{N_M} N_{r,j}$. 

Each coded transmission block spans multiple coherence periods, e.g., multiple distinct resource blocks in an LTE system, of the downlink channel. Specifically, we adopt a block-ergodic channel model, in which the fading channels are constant within a coherence period but vary in an ergodic fashion across a large number of coherence periods. Within each channel coherence period of duration $T$ channel uses, the baseband signal transmitted by the $i$-th RU is given by a $N_{t,i} \times T$ complex matrix ${\bf{X}}_i$, where each column corresponds to the signal transmitted from the $N_{t,i}$ antennas in a channel use.

The $N_{r,j} \times T$ signal ${\bf{Y}}_j$ received by the $j$-th MS in a given channel coherence period, where each column corresponds to the signal received by the $N_{r,j}$ antennas in a channel use, is given by
\begin{equation}
\label{RS;MS} {\bf{Y}}_{j} = {\bf{H}}_j {\bf{X}}  + {\bf{Z}}_{j},
\end{equation}
where ${\bf{Z}}_{j}$ is the $N_{r,j} \times T$ noise matrix, which consist of i.i.d. $\mathcal{CN}(0,1)$ entries; ${\bf{H}}_j = [{\bf{H}}_{j1}, \dots, {\bf{H}}_{j_{N_R}}]$ denotes the $N_{r,j} \times N_t$ channel matrix for $j$-th MS, where ${\bf{H}}_{ji}$ is the $N_{r,j} \times N_{t,i}$ channel matrix from the $i$-th RU to the $j$-th MS; and $ {\bf{X}}$ is the collection of the signals transmitted by all the RUs, i.e., $ {\bf{X}} = [ {\bf{X}}_{1}^T, \dots,  {\bf{X}}_{N_B}^T]^T$. As per the discussion above, the channel matrix ${\bf{H}}_j$ is assumed to be constant during each channel coherence block and to change according to a stationary ergodic process from block to block. We consider both the scenarios in which the CU has either perfect instantaneous information about the channel matrix ${\bf{H}}$ or it is only aware of the distribution of the channel matrix ${\bf{H}}$, i.e., to have {\textit{stochastic CSI}}. Instead, the MSs always have full CSI about their respective channel matrices, as we will state more precisely in the next sections. The transmit signal ${\bf{X}}_i$ has a power constraint given as $E[||{\bf{X}}_i||^2] / T \le \bar P_i$.

\rem \label{CH_Model} A specific channel model of interest is the standard Kronecker model, whereby the channel matrix ${\bf{H}}_{ji}$ is written as 
\begin{equation} \label{ChannelMatrix}
{\bf{H}}_{ji} = {\pmb{\Sigma}}_{R, ji}^{1/2} \widetilde {\bf{H}}_{ji} {\pmb{\Sigma}}_{T, ji}^{1/2},
\end{equation}
where the $N_{t,i} \times N_{t,i}$ matrix ${\pmb{\Sigma}}_{T,ji}$ and the $N_{r,j} \times N_{r,j}$ matrix ${\pmb{\Sigma}}_{R,ji}$ are the transmit-side and receiver-side spatial correlation matrices, respectively, and the $N_{r,j} \times N_{t,i}$ random matrix $\widetilde {\bf{H}}_{ji}$ has i.i.d. $\mathcal{CN}(0,1)$ variables and accounts for the small-scale multipath fading \cite{Caire14TIT}. With this model, stochastic CSI entails that the CU is hence only aware of the correlation matrices ${\pmb{\Sigma}}_{T,ji}$ and ${\pmb{\Sigma}}_{R,ji}$. Moreover, in case that the RUs are placed in a higher location than the MSs, one can assume that the receive-side fading is uncorrelated, i.e., ${\pmb{\Sigma}}_{R,ji} = {\bf{I}}_{N_{r,j}}$, while the transmit-side covariance matrix ${\pmb{\Sigma}}_{T,ji}$ is determined by the one-ring scattering model (see \cite{Caire14TIT} and references therein). In particular, if the RUs are equipped with $\lambda /2$-spaced uniform linear arrays, we have ${\pmb{\Sigma}}_{T,ji} = {\pmb{\Sigma}}_{T}(\theta_{ji}, \Delta_{ji})$ for the $j$-th MS and the $i$-th RU located at a relative angle of arrival $\theta_{ji}$ and having angular spread $\Delta_{ji}$, where the element $(m,n)$ of matrix ${\pmb{\Sigma}}_{T}(\theta_{ji}, \Delta_{ji})$ is given by 
\begin{equation} \label{CorrCH}
[{\pmb{\Sigma}}_{T}(\theta_{ji}, \Delta_{ji})]_{m,n} = \frac{\alpha_{ji}}{2 \Delta_{ji}} \int_{\theta_{ji} - \Delta_{ji}}^{\theta_{ji} + \Delta_{ji}} \exp^{-j \pi (m-n) \sin(\phi)} d\phi,
\end{equation}
with the path loss coefficient $\alpha_{ji}$ between the $j$-th MS and the $i$-th RU being given as 
\begin{equation} \label{PL_coef}
\alpha_{ji} = \frac{1}{1 + \left (\frac{d_{ji}}{d_0}\right )^{\eta}},
\end{equation}
where $d_{ji}$ is the distance between the $j$-th MS and the $i$-th RU, $d_0$ is a reference distance, and $\eta$ is the path loss exponent. $\hspace{15.2cm} \blacksquare$

Each $i$-th fronthaul link has capacity $\bar C_i$, which is measured in bit/s/Hz, where the normalization is with respect to the bandwidth of the downlink channel. In other words, the capacity of the $i$-th fronthaul link is $\bar C_i$ bits per channel use of the downlink. The fronthaul capacity constraint limits the fronthaul rate that is allocated in the coding block, and hence across all the fading states, to be no larger than $\bar C_i$. The fronthaul constraint will be further discussed in Section \ref{Sec:PBC} and \ref{Sec:PBC_CBP}.
\section{Compress-After-Precoding} \label{Sec:PBC}
In this section, we first describe the CAP strategy in Section \ref{Sec:CAP_PF}. Then, we briefly review known strategies for the joint optimization of fronthaul compression and precoding with perfect instantaneous channel knowledge at the CU in Section \ref{Sec:CAP_PerfectCSI}. Finally, we propose an optimization algorithm under the assumption of stochastic CSI at the CU in Section \ref{Sec:CAP_OA}. 

\begin{figure}[t]
\centering
\includegraphics[width=14cm]{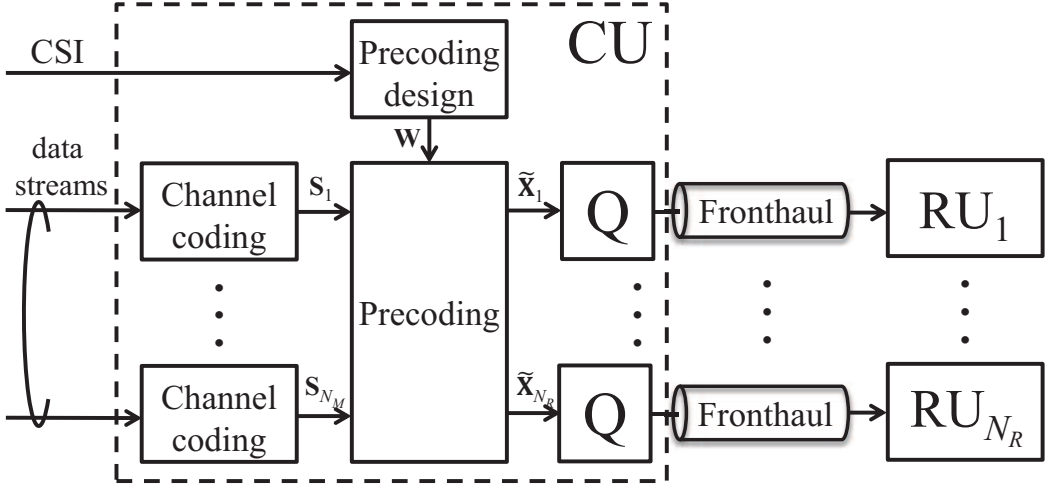}
\caption{Block diagram of the Compression-After-Precoding (CAP) scheme (``${\text{Q}}$" represents fronthaul compression).}
\label{fig:fig2}
\end{figure}
\subsection{Precoding and Fronthaul Compression for CAP} \label{Sec:CAP_PF}
With the CAP scheme as illustrated in Fig. {\ref{fig:fig2}}, the CU performs channel coding and precoding, and then compresses the resulting baseband signals so that they can be forwarded on the fronthaul links to the corresponding RUs. This strategy corresponds to the standard approach envisioned for C-RANs \cite{Simeone09EURADVSP, Marsch09GLOBECOM, Yu14ITA, Park14SPMAG, Park13TSP}. Specifically, channel coding is performed separately for the information stream intended for each MS. This step produces the data signal ${\bf{S}} = [{\bf{S}}_1^\dagger, \dots, {\bf{S}}_{N_M}^\dagger]^\dagger$ for each coherence block, where ${\bf{S}}_j$ is the $M_{j} \times T$ matrix containing, as rows, the $M_j \le N_{r,j}$ encoded data streams for the $j$-th MS. We define the number of total data streams as $M = \sum_{j=1}^{N_M} M_j$ and assume the condition $M \le N_t$. Following standard random coding arguments, we take all the entries of matrix ${\bf{S}}$ to be i.i.d. as $\mathcal{CN}(0,1)$. The encoded data ${\bf{S}}$ is further processed to obtain the transmitted signals ${\bf{X}}$ as detailed below. 

The precoded data signal computed by the CU for any given coherence time can be written as $\widetilde {\bf{X}} = {\bf{W}} {\bf{S}}$, where ${\bf{W}}$ is the $N_t \times M$ precoding matrix. Note that with instantaneous CSI a different precoding matrix ${\bf{W}}$ is used for different coherence times in the coding block, while, with stochastic CSI, the same precoding matrix ${\bf{W}}$ is used for all coherence times. In both cases, the precoded data signal $\widetilde {\bf{X}}$ can be divided into the $N_{t,i} \times T$ signals $\widetilde {\bf{X}}_i$ corresponding to $i$-th RU for all $i \in \mathcal{N}_R$ as $\widetilde {\bf{X}} = [\widetilde {\bf{X}}_1^\dagger, \dots, \widetilde {\bf{X}}_{N_R}^\dagger ]^\dagger$. Specifically, the baseband signal $\widetilde {\bf{X}}_i$ for $i$-th RU is defined as $\widetilde {\bf{X}}_i = {\bf{W}}_i^r {\bf{S}}$, where ${\bf{W}}_i^r$ is the $N_{t,i} \times N_r$ precoding matrix for the $i$-th RU, which is obtained by properly selecting the rows of matrix ${\bf{W}}$ (as indicated by the superscript ``$r$" for ``rows"): the matrix ${\bf{W}}^r_i$ is given as ${\bf{W}}_i^r = {\bf{D}}_i^{r T} {\bf{W}}$, with the $N_t \times N_{t,i}$ matrix ${\bf{D}}_i^r$ having all zero elements except for the rows from $\sum_{k=1}^{i-1} N_{t,k}+1$ to $\sum_{k=1}^i N_{t,k}$, that contain an $N_{t,i} \times N_{t,i}$ identity matrix. 

The CU quantizes each sequence of baseband signal $\widetilde {\bf{X}}_i$ for transmission on the $i$-th fronthaul link to the $i$-th RU. We write the compressed signals $ {\bf{X}}_i$ for $i$-th RU as 
\begin{eqnarray}
\label{PDS;EachRU} {\bf{X}}_i = \widetilde {\bf{X}}_i + {\bf{Q}}_{x,i},
\end{eqnarray}
where the quantization noise matrix ${\bf{Q}}_{x,i}$ is assumed to have i.i.d. $\mathcal{CN}(0, \sigma_{x,i}^2)$ entries. The quantization noises ${\bf{Q}}_{x,i}$ are independent across the RU index $i$, which can be realized via separate quantizers for the signals of different RUs. Note that the possibility to leverage quantization noise correlation across the RUs via joint quantization is explored in \cite{Park14SPMAG, Park13TSP} for static channels. Based on (\ref{PDS;EachRU}), the design of the fronthaul compression reduces to the optimization of the quantization noise variances $\sigma_{x,1}^2, \dots, \sigma_{x,N_B}^2$. The power transmitted by $i$-th RU  is then computed as 
\begin{equation} \label{PowerConst}
P_i \left({\bf{W}}, \sigma_{x,i}^2 \right) = \frac{1}{T} E  [||{\bf{X}}_i||^2 ] = \textrm{tr} \left( {\bf{D}}_i^{rT} {\bf{W}} {\bf{W}}^\dagger {\bf{D}}_i^r + \sigma_{x,i}^2 {\bf{I}} \right), 
\end{equation}
where we have emphasized the dependence of the power $P_i ({\bf{W}}, \sigma_{x,i}^2 )$ on the precoding matrix ${\bf{W}}$ and quantization noise variances $\sigma_{x,i}^2$. Moreover, using standard rate-distortion arguments, the rate required on the fronthaul between the CU and $i$-th RU in a given coherence interval can be quantified by $I (\widetilde {\bf{X}}_i; {\bf{X}}_i ) / T$ (see, e.g., \cite[Ch. 3]{GamalBook}). Therefore, the rate allocated on the $i$-th fronthaul link is equal to
\begin{equation} \label{BC;CAP}
C_i \left({\bf{W}}, \sigma_{x,i}^2 \right ) =  \log \det \left( {\bf{D}}_i^{rT} {\bf{W}} {\bf{W}}^\dagger {\bf{D}}_i^r+ \sigma_{x,i}^2 {\bf{I}} \right) - N_{t,i} \log \left( \sigma_{x,i}^2 \right),
\end{equation}
so that the fronthaul capacity constraint is $C_i ({\bf{W}}, \sigma_{x,i}^2 ) \le \bar C_i$.

We assume that each $j$-th MS is aware of the effective receive channel matrices $\widetilde {\bf{H}}_{jk} = {\bf{H}}_j {\bf{W}}_k^c$ for all $k \in \mathcal{N}_M$ at all coherence times, where ${\bf{W}}^c_k$ is the $N_t \times N_{r,j}$ precoding matrix corresponding to $k$-th MS, which is obtained from the precoding matrix ${\bf{W}}$ by properly selecting the columns as ${\bf{W}} = [{\bf{W}}^c_1, \dots, {\bf{W}}^c_{N_M}]$. We collect the effective channels in the matrix $\widetilde {\bf{H}}_j = [\widetilde{\bf{H}}_{j1}, \dots, \widetilde{\bf{H}}_{j N_M}] = {\bf{H}}_j {\bf{W}}$. The effective channel $\widetilde {\bf{H}}_j$ can be estimated at the MSs via downlink training. Under this assumption, the ergodic achievable rate for the $j$-th MS is computed as $E [R_j^{CAP} ({\bf{H}}, {\bf{W}}, {\pmb{\sigma}}_{x}^2 )]$, with $R_j^{CAP} ({\bf{H}}, {\bf{W}}, {\pmb{\sigma}}_{x}^2 ) = I_{{\bf{H}}} ({\bf{S}}_j; {\bf{Y}}_j )/T$, where $I_{{\bf{H}}} (\widetilde {\bf{S}}_j; {\bf{Y}}_j )$ represents the mutual information conditioned on the value of channel matrix ${\bf{H}}$, the expectation is taken with respect to ${\bf{H}}$ and 
\begin{equation} \label{EAR_CAP}
R_j^{CAP} \hspace{-0.15cm} \left({\bf{H}}, {\bf{W}}, {\pmb{\sigma}}_{x}^2 \right) \hspace{-0.1cm} =  \log \det \left( \hspace{-0.05cm} {\bf{I}} \hspace{-0.05cm} + \hspace{-0.05cm} {\bf{H}}_j \left( {\bf{W}} {\bf{W}}^\dagger \hspace{-0.1cm} + {\bf{\Omega}}_x \right) {\bf{H}}_j^\dagger \right) - \log \det \hspace{-0.1cm} \left ( \hspace{-0.1cm} {\bf{I}} \hspace{-0.05cm} + \hspace{-0.05cm} {\bf{H}}_j \left( \sum_{k \in \mathcal{N}_M \setminus j} \hspace{-0.1cm} {\bf{W}}^c_k {{\bf{W}}_k^c} ^\dagger \hspace{-0.05cm} +\hspace{-0.05cm}  {\bf{\Omega}}_x \right) {\bf{H}}_j^\dagger \right),
\end{equation}
with the covariance matrix ${\bf{\Omega}}_x$ being a diagonal with diagonal blocks given as $\textrm{diag} ([\sigma_{x,1}^2 {\bf{I}}, \dots, \sigma_{x,N_B}^2 {\bf{I}}])$ and ${\pmb{\sigma}}_{x}^2 = [\sigma_{x,1}^2, \dots, \sigma_{x,N_B}^2]^{T}$.

The ergodic achievable weighted sum-rate can be optimized over the precoding matrix ${\bf{W}}$ and the compression noise variances ${\pmb{\sigma}}_{x}^2$ under fronthaul capacity and power constraints. In the next subsections, we consider separately the cases with instantaneous and stochastic CSI. 
\subsection{Perfect Instantaneous CSI} \label{Sec:CAP_PerfectCSI}
In the case of perfect channel knowledge at the CU, the design of the precoding matrix ${\bf{W}}$ and the compression noise variances ${\pmb{\sigma}}_{x}^2$, is adapted to the channel realization ${\bf{H}}$ for each coherence block. To emphasize this fact, we use the notation ${\bf{W}}({\bf{H}})$ and ${\pmb{\sigma}}_{x}^2({\bf{H}})$. The problem of optimizing the ergodic weighted achievable sum-rate with given weights $\mu_j \ge 0$ for $j \in \mathcal{N}_M$ is then formulated as follows:
\begin{subequations} \label{P_CAP_wPerfectCSI:WER_STC}
\begin{eqnarray} 
\underset { {\bf{W}}({\bf{H}}), {\pmb{\sigma}}_{x}^2({\bf{H}}) }{\textrm{maximize}} && \sum_{j\in \mathcal{N}_M} \mu_j E \left[ R_j^{CAP} \left({\bf{H}}, {\bf{W}}({\bf{H}}), {\pmb{\sigma}}_{x}^2({\bf{H}}) \right) \right] \label{OF_CAP_wPerfectCSI:WER_STC} \\
\hspace{0.5cm} \textrm{s.t.} \hspace{0.5cm} && C_i \left({\bf{W}}, \sigma_{x,i}^2({\bf{H}}) \right) \le \bar C_i,  \label{BC_CAP_wPerfectCSI:WER_STC} \\
&& P_i \left({\bf{W}}({\bf{H}}), \sigma_{x,i}^2({\bf{H}}) \right) \le \bar P_i,  \label{PC_CAP_wPerfectCSI:WER_STC}
\end{eqnarray}
\end{subequations}
where (\ref{BC_CAP_wPerfectCSI:WER_STC})-(\ref{PC_CAP_wPerfectCSI:WER_STC}) apply for all $i \in \mathcal{N}_R$ and all channel realizations ${\bf{H}}$. Due to the separability of the fronthaul and power constraints across the channel realizations ${\bf{H}}$, the problem (\ref{P_CAP_wPerfectCSI:WER_STC}) can be solved for each ${\bf{H}}$ independently. Note that the achievable rate in (\ref{OF_CAP_wPerfectCSI:WER_STC}) and the fronthaul constraint in (\ref{BC_CAP_wPerfectCSI:WER_STC}) are non-convex. However, the functions $R_j^{CAP} ({\bf{H}}, {\bf{W}}({\bf{H}}), {\pmb{\sigma}}_{x}^2({\bf{H}}) )$ and $C_i ({\bf{W}}({\bf{H}}), \sigma_{x,i}^2({\bf{H}}) )$ can be then seen to be difference of convex (DC) functions of the covariance matrices $\widetilde {\bf{V}}_j({\bf{H}}) = \widetilde{\bf{W}}_j^c({\bf{H}}) \widetilde{\bf{W}}_j^{c \dagger}({\bf{H}})$ for all $j \in \mathcal{N}_M$ and the variance ${\pmb{\sigma}}_{x}^{2}({\bf{H}})$. The resulting relaxed problem can be tackled via the Majorization-Minimization (MM) algorithm as detailed in \cite{Park14SPMAG, Park13TSP}, from which a feasible solution of problem (\ref{P_CAP_wPerfectCSI:WER_STC}) can be obtained. We refer to \cite{Park14SPMAG, Park13TSP} for details.
\subsection{Stochastic CSI} \label{Sec:CAP_OA}
With only stochastic CSI at the CU, in contrast to the case with instantaneous CSI, the same precoding matrix ${\bf{W}}$ and compression noise variances ${\pmb{\sigma}}_{x}^2$ are used for all the coherence blocks. Accordingly, the problem of optimizing the ergodic weighted achievable sum-rate can be reformulated as follows:
\begin{subequations} \label{P_CAP_woCSI:WER_STC}
\begin{eqnarray} 
\underset {{\bf{W}}, {\pmb{\sigma}}_{x}^2}{\textrm{maximize}} && \sum_{j\in \mathcal{N}_M} \mu_j E \left[ R_j^{CAP} \left({\bf{H}}, {\bf{W}}, {\pmb{\sigma}}_{x}^2 \right) \right] \label{OF_CAP_woCSI:WER_STC} \\
\hspace{0.5cm} \textrm{s.t.} \hspace{0.5cm} && C_i \left({\bf{W}}, \sigma_{x,i}^2 \right) \le \bar C_i,  \label{BC_CAP_woCSI:WER_STC} \\
&& P_i \left({\bf{W}}, \sigma_{x,i}^2 \right) \le \bar P_i,  \label{PC_CAP_woCSI:WER_STC}
\end{eqnarray}
\end{subequations}
where (\ref{BC_CAP_woCSI:WER_STC})-(\ref{PC_CAP_woCSI:WER_STC}) apply to all $i \in \mathcal{N}_R$. In order to tackle this problem, we adopt the Stochastic Successive Upper-bound Minimization (SSUM) method \cite{SSUM_paper}, whereby, at each step, a stochastic lower bound of the objective function is maximized around the current iterate\footnote{We mention here that an alternative method to attack the problem would be the strategy introduced in \cite{Yang13SPAWC}. We leave the study of this approach to future work.}. To this end, similar to \cite{Park14SPMAG, Park13TSP}, we recast the optimization over the covariance matrices ${\bf{V}}_j = {\bf{W}}_j^c {{\bf{W}}_j^c}^\dagger$ for all $j \in \mathcal{N}_M$, instead of the precoding matrices ${\bf{W}}_j^c$ for all $j \in \mathcal{N}_M$. We observe that, with this choice, the objective function is expressed as the average of DC functions, while the constraint (\ref{BC_CAP_woCSI:WER_STC}) is also a DC function, with respect to the covariance ${\bf{V}} = [{\bf{V}}_1 \dots {\bf{V}}_{N_M}]$ and the quantization noise variances ${\pmb{\sigma}}_{x}^2$. As discussed above, the resulting problem is a rank-relaxation of the original problem (\ref{P_CAP_woCSI:WER_STC}). Due to the DC structure, locally tight (stochastic) convex lower bounds can be calculated for objective function (\ref{OF_CAP_woCSI:WER_STC}) and the constraint (\ref{BC_CAP_woCSI:WER_STC}) (see, e.g., \cite{MMBook_tutorial}).

The proposed algorithm based on SSUM \cite{SSUM_paper} contains two nested loops. At each outer iteration $n$, a new channel matrix realization ${\bf{H}}^{(n)} = [{\bf{H}}^{T\,\,(n)}_1, \dots, {\bf{H}}^{T\,\,(n)}_{N_M}]$ is drawn based on the availability of stochastic CSI at the CU. For example, with the model (\ref{ChannelMatrix}), the channel matrices are generated based on the knowledge of the spatial correlation matrices. Following the SSUM scheme, the outer loop aims at maximizing a stochastic lower bound on the objective function, given as 
\begin{equation} \label{CAP_OFwithSSUM}
\frac{1}{n} \sum_{l=1}^{n} \widetilde R_j^{CAP} \left({\bf{H}}^{(l)}, {\bf{V}}, {\pmb{\sigma}}_{x}^{2} | {\bf{V}}^{(l-1)}, {\pmb{\sigma}}_{x}^{2 \,\, (l-1)} \right), 
\end{equation}
where $\widetilde R_j^{CAP}({\bf{H}}^{(l)}, {\bf{V}}, {\pmb{\sigma}}_{x}^{2} | {\bf{V}}^{(l-1)}, {\pmb{\sigma}}_{x}^{2 \,\, (l-1)} )$ is a locally tight convex lower bound on $R_j^{CAP} \left({\bf{H}}, {\bf{W}}, {\pmb{\sigma}}_{x}^2 \right)$ around solution ${\bf{V}}^{(l-1)}$, ${\pmb{\sigma}}_{x}^{2 \,\, (l-1)}$ obtained at the $(l-1)$ the outer iteration when the channel realization is ${\bf{H}}^{(l)}$. This can be calculated as (see, e.g., \cite{SSUM_paper})
\begin{eqnarray} \label{linearizedSR_CAP}
&& \hspace{-0.5cm} \widetilde R_j^{CAP} \left ({\bf{H}}^{(l)}, {\bf{V}}, {\pmb{\sigma}}_{x}^{2} | {\bf{V}}^{(l-1)}, {\pmb{\sigma}}_{x}^{2 \,\, (l-1)} \right) \triangleq \log \det \left( {\bf{I}} + {\bf{H}}_j^{(l)} \left( \sum_{k=1}^{N_M} {\bf{V}}_k + {\bf{\Omega}}_x \right) {\bf{H}}_j^{(l)\,\,\dagger} \right) \\ 
\nonumber && \hspace{1.5cm} - f \left( {\bf{I}} + {\bf{H}}_j^{(l)} \left( \sum_{k=1, k \neq j}^{N_M} {\bf{V}}_k^{(l-1)} + {\bf{\Omega}}_x^{(l-1)} \right) {\bf{H}}_j^{(l)\,\,\dagger}, {\bf{I}} + {\bf{H}}_j^{(l)} \left( \sum_{k=1, k \neq j}^{N_M} {\bf{V}}_k + {\bf{\Omega}}_x \right) {\bf{H}}_j^{(l)\,\,\dagger} \right),
\end{eqnarray}
where the covariance matrix ${\bf{\Omega}}_x^{(l)}$ is a diagonal matrix with diagonal blocks given as $\textrm{diag} ([\sigma_{x,1}^{2 \,\, (l)} {\bf{I}}, \dots, \sigma_{x,N_B}^{2 \,\, (l)} {\bf{I}}])$ and the linearized function $f({\bf{A}}, {\bf{B}})$ is obtained from the first-order Taylor expansion of the log det function as 
\begin{equation} \label{LinearFunc}
f({\bf{A}}, {\bf{B}}) \triangleq \log \det \left( {\bf{A}} \right) + \frac{1}{\textrm{ln} 2} \textrm{tr} \left({\bf{A}}^{-1} \left({\bf{B}} - {\bf{A}} \right) \right).
\end{equation} 
Since the maximization of (\ref{CAP_OFwithSSUM}) is subject to the non-convex DC constraint (\ref{BC_CAP_woCSI:WER_STC}), the inner loop tackles the problem via the MM algorithm i.e., by applying successive locally tight convex lower bounds to the left-hand side of the constraint (\ref{BC_CAP_woCSI:WER_STC}) \cite{MMBook}. Specifically, given the solution ${\bf{V}}^{(n,r-1)}$ and  ${\pmb{\sigma}}_{x}^{2\,\,(n,r-1)}$ at $(r-1)$-th inner iteration of the $n$-th outer iteration, the fronthaul constraint in (\ref{BC_CAP_woCSI:WER_STC}) at the $r$-th inner iteration can be locally approximated as
\begin{eqnarray} \label{linearizedBR_CAP}
&&\hspace{-1cm}\widetilde C_i \left( {\bf{V}}, {{\sigma}}_{x,i}^{2} | {\bf{V}}^{(n,r-1)}, {{\sigma}}_{x,i}^{2 \,\, (n,r-1)} \right) \triangleq \\
\nonumber && \hspace{2cm} f \left( \sum_{k=1}^{N_M} {\bf{D}}_i^{rT} {\bf{V}}_k^{(n,r-1)} {\bf{D}}_i^r + \sigma_{x,i}^{2\,\,(n,r-1)} {\bf{I}}, \sum_{k=1}^{N_M} {\bf{D}}_i^{rT} {\bf{V}}_k {\bf{D}}_i^r + \sigma_{x,i}^{2} {\bf{I}} \right) - N_{t,i} \log \left( \sigma_{x,i}^{2}\right).
\end{eqnarray}
The resulting combination of SSUM and MM algorithms for the solution of problem (\ref{P_CAP_woCSI:WER_STC}) is summarized in Table Algorithm \ref{SSUM}. The algorithm is completed by calculating, from the obtained solution ${\bf{V}}^{*}$ of the relaxed problem, the precoding matrix ${\bf{W}}$ by using the standard rank-reduction approach \cite{BoydSemiProg}, which is given as ${\bf{W}}_j^{*} = \gamma_j {\mathbf{\nu}}_{\textrm{max}}^{(M_j)} ({\bf{V}}_j^{*})$ with the normalization factor $\gamma_j$, selected so as to satisfy the power constraint with equality, namely $P_i \left({\bf{W}}, \sigma_{x,i}^2 \right) = \bar P_i$.

 \linespread{1}
\begin{algorithm} [t]
\begin{algorithmic}
\caption{CAP Design of Fronthaul Compression and Precoding with stochastic CSI} \label{SSUM}
\State {\textbf{Initialization (outer loop)}}: Initialize the covariance matrices ${\bf{V}}^{(0)}$ and the quantization noise variances ${\pmb{\sigma}}_{x}^{2 \,\, (0)}$, and set $n=0$.
\State {\textbf{repeat}} 
\State \indent $n \gets n+1$
\State \indent Generate a channel matrix realization ${\bf{H}}^{(n)}$ using the available stochastic CSI.
\State \indent {\textbf{Initialization (inner loop)}}: Initialize ${\bf{V}}^{(n,0)} = {\bf{V}}^{(n-1)}$ and ${\pmb{\sigma}}_{x}^{2 \,\, (n,0)} = {\pmb{\sigma}}_{x}^{2 \,\, (n-1)}$, and set $r=0$. 
\State \indent {\textbf{repeat}}
\State \indent \indent $r \gets r+1$ 
\begin{eqnarray} \label{ConvexProblem_SSUM_inner}
\hspace{-0.7cm} {\bf{V}}^{(n,r)}, {\pmb{\sigma}}_{x}^{2 \,\, (n,r)} \gets \arg \max_{{{\bf{V}}, {\pmb{\sigma}}_{x}^{2}}} && \frac{1}{n} \sum_{l=1}^{n} \sum_{j\in \mathcal{N}_M} \mu_j \widetilde R_j^{CAP} \left({\bf{H}}^{(l)}, {\bf{V}}, {\pmb{\sigma}}_{x}^{2} | {\bf{V}}^{(l-1)}, {\pmb{\sigma}}_{x}^{2 \,\, (l-1)} \right) \\
\nonumber {\textrm{s.t.}} \hspace{0.1cm} && \widetilde C_i \left({\bf{V}}, {{\sigma}}_{x,i}^{2} | {\bf{V}}^{(n,r-1)}, {{\sigma}}_{x,i}^{2 \,\, (n,r-1)} \right) \le \bar C_i, \\
\nonumber && P_i \left ({\bf{V}}, \sigma_{x,i}^{2} \right) \le \bar P_i,  \hspace{0.3cm} \textrm{for all} \,\, i \in \mathcal{N}_R.
\end{eqnarray}
\State \indent {\textbf{until}} a convergence criterion is satisfied.
\State \indent Update ${\bf{V}}^{(n)} \gets {\bf{V}}^{(n,r)}$ and ${\pmb{\sigma}}_{x}^{2 \,\, (n)} \gets {\pmb{\sigma}}_{x}^{2 \,\, (n,r)}$
\State {\textbf{until}} a convergence criterion is satisfied. 
\State {\textbf{Solution}}: Calculate the precoding matrix ${\bf{W}}$ from the covariance matrices ${\bf{V}}^{(n)}$ via rank reduction as ${\bf{W}}_j = \gamma_j  {\mathbf{\nu}}_{\textrm{max}}^{(M_j)} ({\bf{V}}_j^{(n)})$ for all $j \in \mathcal{N}_M$, where $\gamma_j$ is obtained by imposing $P_i \left({\bf{W}}, \sigma_{x,i}^2 \right) = \bar P_i$ using (\ref{PowerConst}).
\end{algorithmic}
\end{algorithm}
\linespread{2} 

Two remarks are in place on the properties of the proposed algorithm. First, since the approximated functions (\ref{linearizedSR_CAP}) and (\ref{linearizedBR_CAP}) are local lower bounds, the algorithm provides a feasible solution of the relaxed problem at each inner and outer iteration (see, e.g., \cite{SSUM_paper}). The second remark is that, from \cite{SSUM_paper, MMBook_tutorial}, as long as a sufficient number of inner iterations is performed at each outer iteration, the algorithm is guaranteed to converge to stationary points of the relaxed problem. 
\section{Compression-Before-Precoding} \label{Sec:PBC_CBP}
\begin{figure}[t]
\centering
\includegraphics[width=17cm]{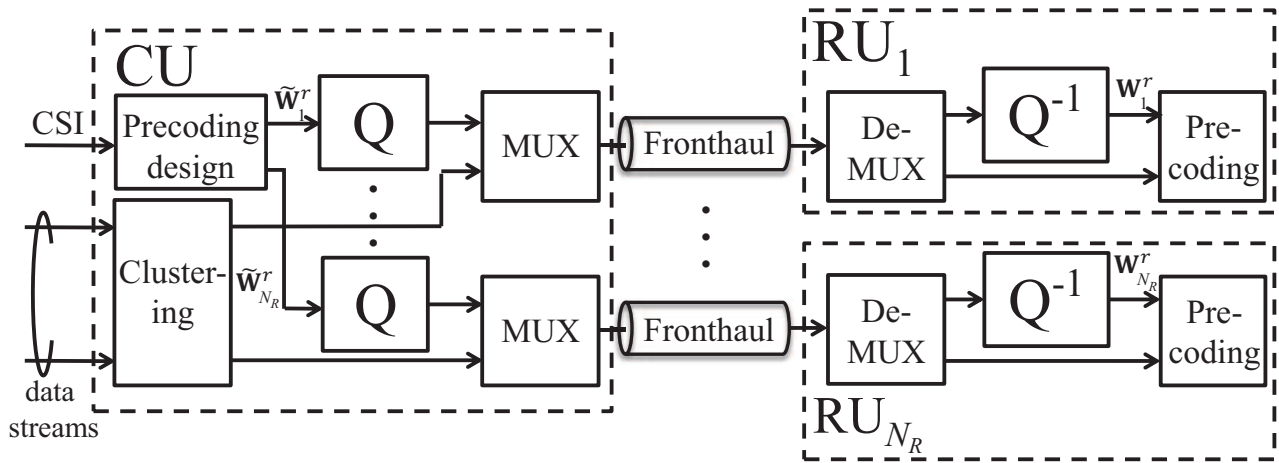}
\caption{Block diagram of the Compression-Before-Precoding (CBP) scheme (``${\text{Q}}$" and ``${\text{Q}}^{-1}$" represents fronthaul compression and decompression, respectively).}
\label{fig:fig3}
\end{figure}

With the Compression-Before-Precoding (CBP) scheme, the CU calculates the precoding matrices, but does not perform precoding. Instead, as illustrated in Fig. {\ref{fig:fig3}}, it uses the fronthaul links to communicate the information messages of a given subset of MSs to each RU, along with the corresponding compressed precoding matrices. Each RU can then encode and precode the messages of the given MSs based on the information received from the fronthaul link. As it will be discussed, in the CBP scheme, unlike CAP, a preliminary clustering step is generally advantageous whereby each MS is assigned to a subset of RUs. In the following, we first describe the CBP strategy in Section \ref{CBP_PF}; then we review the design problem under instantaneous CSI in Section \ref{Sec:CBP_PerfectCSI}; and, finally, we introduce an algorithm for the joint optimization of fronthaul compression and precoding with stochastic CSI at the CU.

\subsection{Precoding and Fronthaul Compression for CBP} \label{CBP_PF}
As shown in Fig. \ref{fig:fig3}, in the CBP method, the precoding matrix $\widetilde {\bf{W}}$ and the information streams are separately transmitted from the CU to the RUs, and the received information bits are encoded and precoded at each RU using the received precoding matrix. Note that, with this scheme, the transmission overhead over the fronthaul depends on the number of MSs supported by a RU, since the RUs should receive all the corresponding information streams.

Given the above, with the CBP strategy, we allow for a preliminary clustering step at the CU whereby each RU is assigned by a subset of the MSs. We denote the set of MSs assigned by $i$-th RU as $\mathcal{M}_i \subseteq \mathcal{N}_M$ for all $i \in \mathcal{N}_B$. This implies that $i$-th RU only needs the information streams intended for the MSs in the set $\mathcal{M}_i$. We also denote the set of RUs that serve the $j$-th MS, as $\mathcal{B}_j = \{i|j\in \mathcal{M}_i\} \subseteq \mathcal{N}_B$ for all $j \in \mathcal{N}_M$. We use the notation $\mathcal{M}_i[k]$ and $\mathcal{B}_j[m]$ to respectively denote the $k$-th MS and $m$-th RU in the sets $\mathcal{M}_i$ and $\mathcal{B}_j$, respectively. We define the number of all transmit antennas for the RUs, which serve the $j$-th MS, as $N_{t, \mathcal{B}_j}$. We assume here that the sets of MSs assigned by $i$-th RU are given and not subject to optimization (see Section \ref{Sec:Numerical Results} for further details).

The precoding matrix $\widetilde {\bf{W}}$ is constrained to have zeros in the positions that correspond to RU-MS pairs such that the MS is not served by the given RU. This constraint can be represented as 
\begin{eqnarray}
\widetilde {\bf{W}} = \left[{\bf{E}}_1^c \widetilde {\bf{W}}_1^c, \dots, {\bf{E}}_{N_M}^c \widetilde {\bf{W}}_{N_M}^c \right],
\end{eqnarray}
where $\widetilde {\bf{W}}_j^c$ is the $N_{t, \mathcal{B}_j} \times N_{r,j}$ precoding matrix intended for $j$-th MS and RUs in the cluster $\mathcal{B}_j$, and the $N_t \times N_{t, \mathcal{B}_j}$ constant matrix ${\bf{E}}_j^c$ (${\bf{E}}_j^c$ only has either a 0 or 1 entries) defines the association between the RUs and the MSs as ${\bf{E}}_j^c = \left[ {\bf{D}}_{\mathcal{B}_j[1]}^c, \dots, {\bf{D}}_{\mathcal{B}_j[|\mathcal{B}_j|]}^c \right]$, with the $N_r \times N_{r,j}$ matrix ${\bf{D}}_j^c$ having all zero elements except for the rows from $\sum_{k=1}^{j-1} N_{r,k}+1$ to $\sum_{k=1}^j N_{r,j}$, which contain an $N_{r,j} \times N_{r,j}$ identity matrix. 

The sequence of the $N_{t,i} \times N_{r,\mathcal{M}_i}$ precoding matrices $\widetilde {\bf{W}}_i^r$ intended for each $i$-th RU for all coherence times in the coding block is compressed by the CU and forwarded over the fronthaul link to the $i$-th RU. The compressed precoding matrix ${\bf{W}}^r_i$ for $i$-th RU is given by 
\begin{eqnarray}
\label{PM;EachRU} {\bf{W}}_i^r = \widetilde {\bf{W}}_i^r + {\bf{Q}}_{w,i},
\end{eqnarray}
where the $N_{t,i} \times N_{r, \mathcal{M}_i}$ quantization noise matrix ${\bf{Q}}_{w,i}$ is assumed to have zero-mean i.i.d. $\mathcal{CN}(0, \sigma_{w,i}^2 )$ entries and to be independent across the index $i$. Overall, the $N_t \times N_r$ compressed precoding matrix ${\bf{W}}$ for all RUs is represented as
\begin{eqnarray}
\label{PM;AllRU} {\bf{W}} = \widetilde {\bf{W}} + {\bf{Q}}_{w},
\end{eqnarray}
where ${\bf{W}} = [{\bf{E}}_1^{r \dagger} {\bf{W}}_{w,1}^\dagger, \dots,  {\bf{E}}_{N_B}^{r \dagger} {\bf{W}}_{w,N_B}^\dagger]^\dagger$, $\widetilde {\bf{W}}$ and ${\bf{Q}}_w$ are similarly defined. Note that we have $E [\textrm{vec}({\bf{Q}}_{w})$ $\textrm{vec}({\bf{Q}}_{w})^\dagger ] = {\bf{\Omega}}_{w}$, where ${\bf{\Omega}}_{w}$ is a diagonal matrix with diagonal blocks given by $[\sigma_{w,1}^2 {\bf{I}}, \dots, \sigma_{w,N_B}^2 {\bf{I}}]$. 

The ergodic rate achievable for $j$-th MS can be written as $E[ R_j^{CBP} ({\bf{H}}, \widetilde {\bf{W}}, {\pmb{\sigma}}_{w}^2 )]$, where 
\begin{eqnarray} \label{ARC;CBP}
&& \hspace{-0.7cm} R_j^{CBP}\left ({\bf{H}}, \widetilde {\bf{W}}, {\pmb{\sigma}}_{w}^2 \right) = \frac{1}{T} I_{{\bf{H}}} \left({\bf{S}}_j; {\bf{Y}}_j  \right) = \log \det \left( {\bf{I}} + {\bf{H}}_j \left(\widetilde {\bf{W}} \widetilde {\bf{W}}^\dagger + {\bf{\Omega}}_w \right) {\bf{H}}_j^\dagger \right) \\ 
\nonumber && \hspace{7.5cm} - \log \det \left ( {\bf{I}} + {\bf{H}}_j \left( \sum_{k \in \mathcal{N}_M \setminus j} \widetilde {\bf{W}}_k^c \widetilde {\bf{W}}_k^{c \dagger} + {\bf{\Omega}}_w \right) {\bf{H}}_j^\dagger \right).
\end{eqnarray}

\subsection{Perfect Instantaneous CSI} \label{Sec:CBP_PerfectCSI}
With perfect CSI at the CU, as discussed in Section \ref{Sec:CAP_PerfectCSI}, one can adopt the precoding matrix $\widetilde {\bf{W}}({\bf{H}})$, the user rates $\{R_j({\bf{H}})\}$ and the quantization noise variances ${\pmb{\sigma}}_{w}^2({\bf{H}})$ to the current channel realization at each coherence block. The rate required to transmit precoding information on the $i$-th fronthaul in a given channel realizations ${\bf{H}}$ is given by $C_i ({\bf{H}}, \widetilde {\bf{W}}_i^r , \sigma_{w,i}^2 )/T$, with 
\begin{equation} \label{BC;CBP}
\frac{1}{T}C_i\left({\bf{H}}, \widetilde {\bf{W}}_i^r , \sigma_{w,i}^2 \right) = \frac{1}{T} I_{\bf{H}} (\widetilde {\bf{W}}_i^r; {\bf{W}}_i^r ) = \frac{1}{T} \left\{ \log \det \left( {\bf{D}}_i^{rT} \widetilde {\bf{W}} \widetilde {\bf{W}}^\dagger {\bf{D}}_i^r + \sigma_{w,i}^2 {\bf{I}}  \right) - N_{t,i} \log \left( \sigma_{w,i}^2 \right) \right\},
\end{equation}
where the rate $C_i(\widetilde {\bf{W}}_i^r, \sigma_{w,i}^2 )$ required on $i$-fronthaul link is defined in (\ref{BC;CAP}). Note that the normalization by $T$ is needed since only a single precoding matrix is needed for each channel coherence interval. Then, under the fronthaul capacity constraint, the remaining fronthaul capacity that can be used to convey precoding information corresponding to the $i$-th RU is $\bar C_i - \sum_{j \in \mathcal{M}_i} R_j$. As a result, the optimization problem of interest can be formulated as 
\begin{subequations} \label{P_CBP_wPerfectCSI;WER_LTC}
\begin{eqnarray}
\label{OF_CBP_wPerfectCSI;WER_LTC} \hspace{-1cm} \underset {\widetilde {\bf{W}}({\bf{H}}), \, {\pmb{\sigma}}^2_{w,i}({\bf{H}}), \{ R_j({\bf{H}})\} }{\textrm{maximize}} &&  \sum_{j\in \mathcal{N}_M} \mu_j R_j({\bf{H}})   \\
\label{RC_CBP_wPerfectCSI;WER_LTC} \hspace{0.5cm} s.t. \hspace{0.7cm} && R_j({\bf{H}}) \le R_j^{CBP} \left({\bf{H}}, \widetilde {\bf{W}}({\bf{H}}), {\pmb{\sigma}}^2_{w}({\bf{H}}) \right), \\
\label{BC_CBP_wPerfectCSI;WER_LTC} \hspace{-1cm} && \frac{1}{T} C_i \left({\bf{H}}, \widetilde {\bf{W}}_i^r({\bf{H}}) , \sigma_{w,i}^2({\bf{H}}) \right) \le \bar C_i - \sum_{j \in \mathcal{M}_i} R_j({\bf{H}}), \\
\label{PC_CBP_wPerfectCSI;WER_LTC} \hspace{-1cm} && P_i \left (\widetilde {\bf{W}}_i^r({\bf{H}}), \sigma_{w,i}^2({\bf{H}}) \right) \le \bar P_i,  
\end{eqnarray}
\end{subequations}
where the constraints apply to all channel realization, (\ref{RC_CBP_wPerfectCSI;WER_LTC}) applies to all $j \in \mathcal{N}_M$, (\ref{BC_CBP_wPerfectCSI;WER_LTC}) - (\ref{PC_CBP_wPerfectCSI;WER_LTC}) apply to all $i \in \mathcal{N}_R$ and the transmit power $P_i (\widetilde {\bf{W}}_i^r({\bf{H}}), \sigma_{w,i}^2({\bf{H}}))$ at $i$-th RU is defined in (\ref{PowerConst}). Similar to Section \ref{Sec:CAP_PerfectCSI}, the problem (\ref{P_CBP_wPerfectCSI;WER_LTC}) can be studied  for each ${\bf{H}}$ independently. In addition, each subproblem can be tackled by using MM algorithm as explained in \cite{Park14SPMAG, Park13TSP}.

\subsection{Stochastic CSI} \label{Sec:CBP_Stochastic}
\linespread{1}
\begin{algorithm} [t]
\caption{CBP Design of Fronthaul Compression and Precoding with stochastic CSI} \label{SSUM_CBP}
\begin{algorithmic}
\State {\textbf{Initialization}}: Initialize the covariance matrices $\widetilde {\bf{V}}^{(0)}$ and the user rate $\{R_j^{(0)}\}$ and set $n=0$.
\State {\textbf{repeat}}
\State \indent $n \gets n+1$ 
\State \indent Generate a channel matrix realization ${\bf{H}}^{(n)}$ using the available stochastic CSI.
\begin{eqnarray} \label{ConvexProblem_SSUM_CBP_inner}
\hspace{-1cm} \widetilde {\bf{V}}^{(n)}, \{R_j^{(n)}\} \gets \arg \max_{\widetilde {\bf{V}}, \{ R_j\}} &&\hspace{-0.5cm} \sum_{j\in \mathcal{N}_M} \mu_j R_j \\
\nonumber  {\textrm{s.t.}} \hspace{0.4cm} && \hspace{-0.5cm} R_j \le \frac{1}{n} \sum_{l=1}^{n} \widetilde R_j^{CBP} \left({\bf{H}}^{(l)}, \widetilde {\bf{V}} |\widetilde  {\bf{V}}^{(l-1)} \right), \\
\nonumber \hspace{-1cm} &&\hspace{-0.5cm}  \sum_{j \in \mathcal{M}_i} R_j \le \bar C_i, \\
\nonumber \hspace{-1cm} &&\hspace{-0.5cm} P_i \left( \widetilde {\bf{V}}, 0 \right) \le \bar P_i, \hspace{0.5cm} \textrm{for all} \,\, i \in \mathcal{N}_R \,\, \textrm{and} \,\, j \in \mathcal{N}_M.
\end{eqnarray}
\State {\textbf{until}} a convergence criterion is satisfied. 
\State {\textbf{Solution}}: Calculate the precoding matrix $\widetilde {\bf{W}}$ from the covariance matrices $\widetilde{\bf{V}}^{(n)}$ via rank reduction as $\widetilde{\bf{W}}_j = \gamma_j  {\mathbf{\nu}}_{\textrm{max}}^{(M_j)} (\widetilde {\bf{V}}_j^{(n)})$ for all $j \in \mathcal{N}_M$, where $\gamma_j$ is obtained by imposing $P_i \left(\widetilde {\bf{W}} \right) = \bar P_i$ using (\ref{PowerConst}).
\end{algorithmic}
\end{algorithm}
\linespread{2}
With stochastic CSI at the CU, the same precoding matrix is used for all the coherence blocks and hence the rate required to convey the precoding matrix $\widetilde {\bf{W}}_i^r$ to each $i$-th RU becomes negligible. As a result, we can neglect the effect of the quantization noise and set $\sigma_{w,i}^2 = 0$ for all $i \in \mathcal{N}_R$. Accordingly, the fronthaul capacity can be only used for transfer of the information stream as $ \sum_{j \in \mathcal{M}_i} R_j \le C_i$, for all $i \in \mathcal{N}_R$. Based on the above considerations, the optimization problem of interest is formulated as 
\begin{subequations} \label{P_CBP_woCSI;WER_LTC}
\begin{eqnarray}
\label{OF_CBP_woCSI;WER_LTC} \hspace{-1cm} \underset {\widetilde {\bf{W}}, \{ R_j\} }{\textrm{maximize}} &&  \sum_{j\in \mathcal{N}_M} \mu_j R_j   \\
\label{RC_CBP_woCSI;WER_LTC} \hspace{0.5cm} s.t. \hspace{0.7cm} && R_j \le E \left [ R_j^{CBP} \left({\bf{H}}, \widetilde {\bf{W}}, {\pmb{0}} \right) \right], \\
\label{BC_CBP_woCSI;WER_LTC} \hspace{-1cm} && \sum_{j \in \mathcal{M}_i} R_j \le \bar C_i, \\
\label{PC_CBP_woCSI;WER_LTC} \hspace{-1cm} && P_i \left (\widetilde {\bf{W}}_i^r, 0 \right) \le \bar P_i,  
\end{eqnarray}
\end{subequations}
where (\ref{RC_CBP_woCSI;WER_LTC}) applies to all $j \in \mathcal{N}_M$, (\ref{BC_CBP_woCSI;WER_LTC})-(\ref{PC_CBP_woCSI;WER_LTC}) apply to all $i \in \mathcal{N}_R$ and the transmit power $P_i (\widetilde {\bf{W}}_i^r, \sigma_{w,i}^2)$ at $i$-th RU is defined in (\ref{PowerConst}). In problem (\ref{P_CBP_woCSI;WER_LTC}), the constraint (\ref{RC_CBP_woCSI;WER_LTC}) is not only non-convex but also stochastic. Similar to Section \ref{Sec:CAP_OA}, the functions $R_j^{CBP}({\bf{H}}, \widetilde {\bf{W}})$ can be seen to be DC functions of the covariance matrices $\widetilde {\bf{V}}_j = \widetilde{\bf{W}}_j^c \widetilde{\bf{W}}_j^{c \dagger}$ for all $j \in \mathcal{N}_M$, hence opening up the possibility to develop a solution based on SSUM. Referring to Section \ref{Sec:CAP_OA}, for details, given the solutions $\widetilde  {\bf{V}}^{(l-1)}$ at the previous iterations, $l \le n$, the algorithm approximates the function $E[R_j^{CBP}({\bf{H}}, \widetilde {\bf{W}})]$ in (\ref{RC_CBP_woCSI;WER_LTC}) with the stochastic upper bound as 
\begin{equation}
\frac{1}{n} \sum_{l=1}^{n} \widetilde R_j^{CBP} \left({\bf{H}}^{(l)}, \widetilde {\bf{V}} | \widetilde {\bf{V}}^{(l-1)} \right)
\end{equation}
with
\begin{eqnarray} \label{LinearizedRate_CBP}
&& \hspace{-1cm} \widetilde R_j^{CBP} \left(\widetilde  {\bf{V}} |{\bf{H}}^{(l)}, \widetilde  {\bf{V}}^{(l-1)} \right) \triangleq \log \det \left( {\bf{I}} + {\bf{H}}_j^{(l)} \left( \sum_{k=1}^{N_M} \widetilde {\bf{V}}_k \right) {\bf{H}}_j^{\dagger\,\,(l)} \right) \\ 
\nonumber && \hspace{4cm} - f \left( {\bf{I}} + {\bf{H}}_j^{(l)} \left( \sum_{k=1, k \neq j}^{N_M} \widetilde  {\bf{V}}_k^{(l-1)} \right) {\bf{H}}_j^{\dagger\,\,(l)}, {\bf{I}} + {\bf{H}}_j^{(l)} \left( \sum_{k=1, k \neq j}^{N_M} \widetilde  {\bf{V}}_k \right) {\bf{H}}_j^{\dagger\,\,(l)} \right),
\end{eqnarray}
where the linearization function $f({\bf{A}},{\bf{B}})$ is defined in (\ref{LinearFunc}). The algorithm which is summarized in Table Algorithm \ref{SSUM_CBP}, has the same properties discussed for the algorithm in Table Algorithm \ref{SSUM}, namely it provides a feasible solution of the relaxed problem at each iteration and it converge to a stationary point of the same problem. 
\section{Numerical Results} \label{Sec:Numerical Results}
\begin{figure}[t]
\centering
\includegraphics[width=11cm]{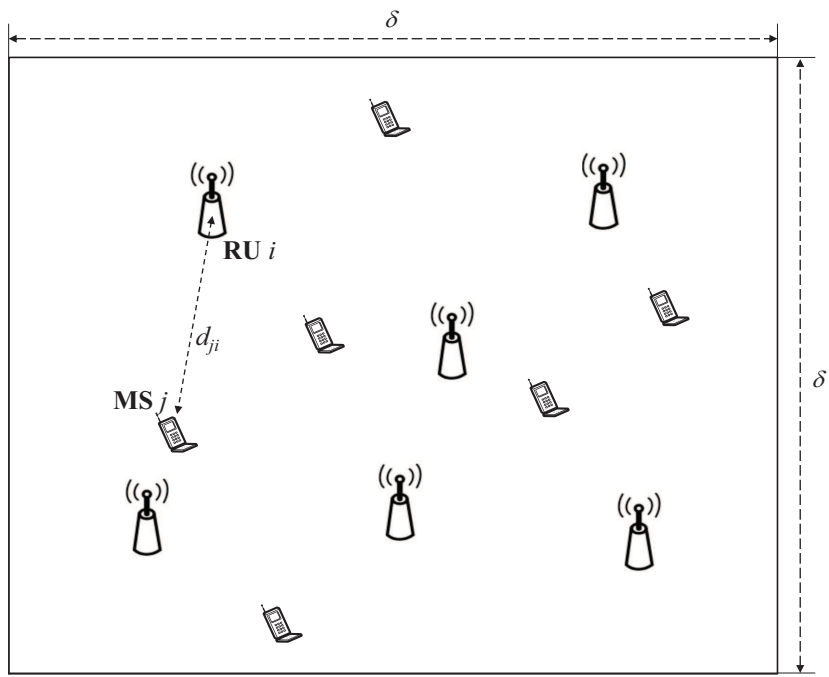}
\caption{Set-up under consideration for the numerical results in Section \ref{Sec:Numerical Results}, where the RUs are randomly located in a square with side $\delta$ and all MSs and RUS are randomly uniformly placed.}
\label{Fig:SimulEnviron}
\end{figure}
In this section, we compare the performance of the CAP and CBP schemes in the set-up under study of block-ergodic channels. To this end, we consider a system in which the RUs and the MSs are randomly located in a square area with side $\delta=500m$ as in Fig. \ref{Fig:SimulEnviron}. In the path loss formula (\ref{PL_coef}), we set the reference distance to $d_0=50m$ and the path loss exponent to $\eta = 3$. We adopt the spatial correlation model in (\ref{CorrCH}) with the angular spread $\Delta_{ji} = \arctan (r_s/d_{ji})$, with the scattering radius $r_s = 10m$ and with $d_{ji}$ being the Euclidean distance between the $i$-th RU and the $j$-th MS. Throughout, we assume that the every RU is subject to the same power constraint $\bar P$ and has the same fronthaul capacity $\bar C$, that is $\bar P_i = \bar  P$ and $\bar C_i = \bar C$ for $i \in \mathcal{N}_R$. Moreover, in the CBP scheme, the MS-to-RU assignment is carried out by choosing, for each RU, the $N_c$ MSs that have the largest instantaneous channel norms for instantaneous CSI and the largest average channel matrix norms for stochastic CSI. Note that this assignment is done for each coherence block in the former case, while in the latter the same assignment holds for all coherence blocks. Note also that a given MS is generally assigned to multiple RUs. 
\begin{figure}[t]
\centering
\includegraphics[width=13cm]{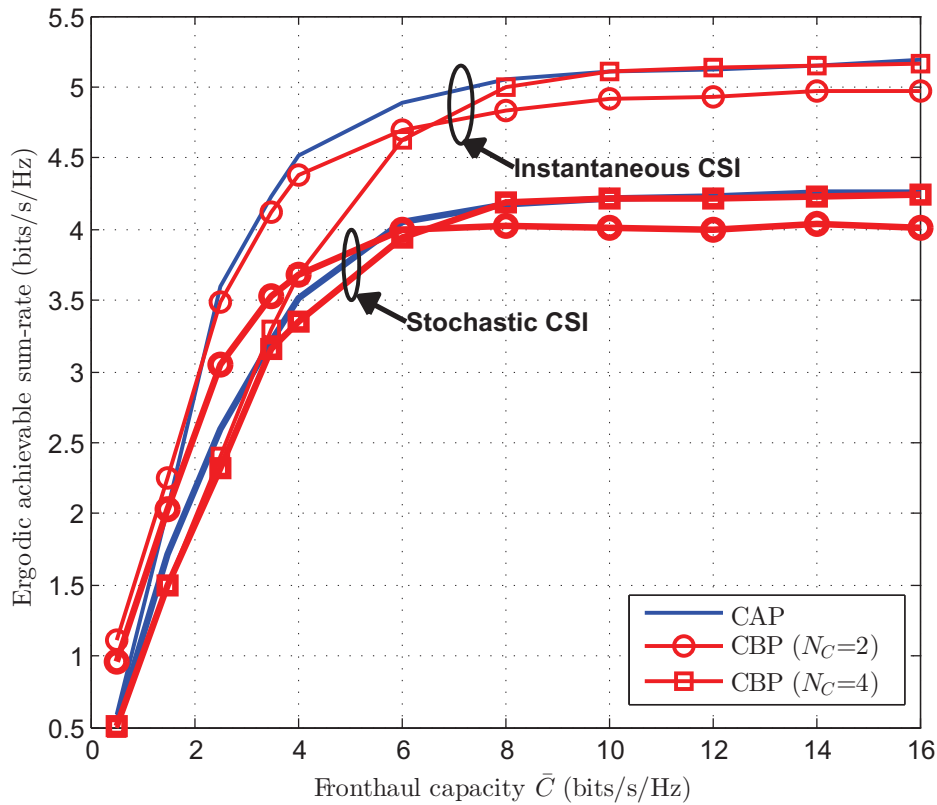}
\caption{Ergodic achievable sum-rate vs. the fronthaul capacity $\bar C$ ($N_R = N_M = 4$, $N_{t,i} = 2$, $N_{r,j}=1$, $\bar P=10$ dB, $T = 20$, and $\mu=1$).}
\label{Fig:CAPandCBP_P0T20NR4}
\end{figure}


The effect of the fronthaul capacity limitation on the ergodic achievable sum-rate is investigated in Fig. \ref{Fig:CAPandCBP_P0T20NR4}, where the number of RUs and MSs is $N_R = N_M = 4$, the number of transmit antennas is $N_{t,i} = 2$ for all $i \in \mathcal{N}_R$, the number of receive antennas is $N_{r,j}=1$ for all $j \in \mathcal{N}_M$, the power is $\bar P=10dB$, and the coherence time is $T = 20$. We first observe that, with instantaneous CSI, the CAP strategy is uniformly better than CBP as long as the fronthaul capacity is sufficiently large (here $\bar C > 2$). This is due to the enhanced interference mitigation capabilities of CAP resulting from its ability to coordinate all the RUs via joint baseband processing without requiring the transmission of all messages on all fronthaul links. Note, in fact, that, with CBP, only $N_c$ MSs are served by each RU, and that making $N_c$ larger entails a significant increase in the fronthaul capacity requirements. We will later see that this advantage of CAP is offset by the higher fronthaul efficiency of CBP in transmitting precoding information for large coherence periods $T$ (see Fig. \ref{Fig:CAPandCBP_C8P0NR4}). Instead, with stochastic CSI, in the low fronthaul capacity regime, here about $\bar C < 6$, the CBP strategy is generally advantageous due to the additional advantage that is accrued by amortizing the precoding overhead over the entire coding block. Another observation is that, for small $\bar C$, the CBP schemes with progressively smaller $N_c$ have better performance thanks to the reduced fronthaul overhead. Moreover, for large $\bar C$, the performance of the CBP scheme with $N_c=N_M$, whereby each RU serves all MSs, approaches that of the CAP scheme. 

\begin{figure}[t]
\centering
\includegraphics[width=13cm]{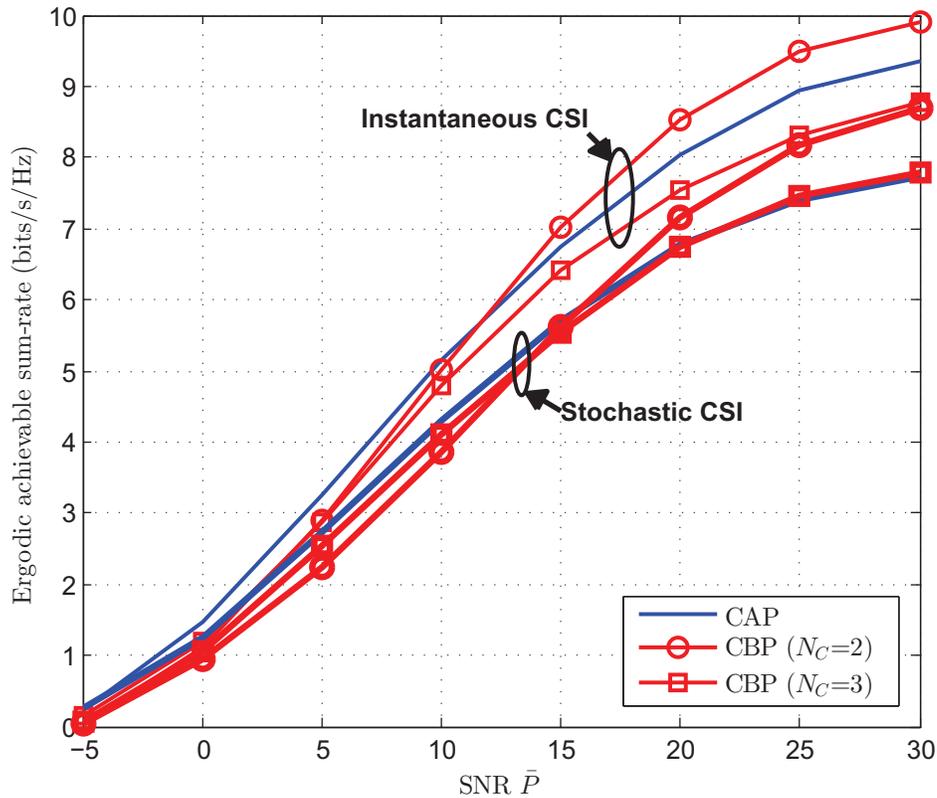}
\caption{Ergodic achievable sum-rate vs. the power constraint $\bar P$ ($N_R = N_M = 4$, $N_{t,i} = 2$, $N_{r,j}=1$, $\bar C=6$ bits/s/Hz, $T = 15$, and $\mu=1$).}
\label{Fig:CAPandCBP_C6T15NR4}
\end{figure}

\begin{figure}[h!]
\centering
\includegraphics[width=13cm]{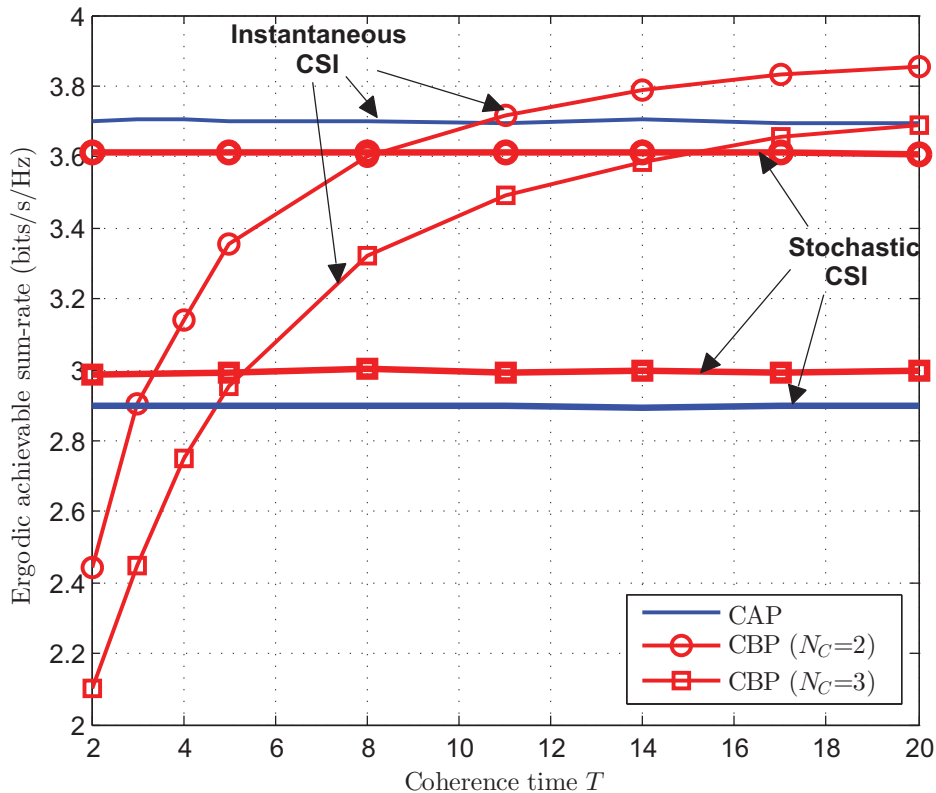}
\vspace{-0.5cm}
\caption{Ergodic achievable sum-rate vs. the coherence time $T$ ($N_R = N_M = 4$, $N_{t,i} = 2$, $N_{r,j}=1$, $\bar C = 2$ bits/s/Hz, $\bar P=20dB$, and $\mu=1$).}
\label{Fig:CAPandCBP_C8P0NR4}
\end{figure}


The effect of the power constraint $\bar P$ is investigated in Fig. \ref{Fig:CAPandCBP_C6T15NR4}, where the number of RUs and MSs is $N_R = N_M = 4$, the number of transmit antennas is $N_{t,i} = 2$, the number of receive antennas is $N_{r,j}=1$, the fronthaul capacity is $\bar C = 6$ bits/s/Hz, and the coherence time is $T = 15$. As a general rule, increasing $\bar P$ enhances the relative impact of the quantization noise on the performance. This can be seen from, e.g., (\ref{BC;CAP}), from which it follows that the quantization noise variance increases with the power $\bar P$ for a fixed value of the fronthaul capacity $\bar C$. The CAP approach is seen to be advantageous in the low power regime, in which the RU coordination gains are not offset by the effect of the quantization noise. In contrast, the CBP method is to be preferred in the larger power regime due to the limited impact of the quantization noise on its performance since only precoding information is quantized. 

Fig. \ref{Fig:CAPandCBP_C8P0NR4} shows the ergodic achievable sum-rate as function of the coherence time $T$, with $N_R = N_M = 4$,  $N_{t,i} = 2$, $N_{r,j}=1$, $\bar C = 2$ bits/s/Hz, and $\bar P=20$ dB. As anticipated, with instantaneous CSI, CBP is seen to benefit from a larger coherence time $T$, since the fronthaul overhead required to transmit precoding information gets amortized over a larger period. This is in contrast to CAP for which such overhead scales proportionally to the coherence time $T$ and hence the CAP scheme is not affected by the coherence time. As a result, CBP can outperform CAP for sufficiently large $T$ in the presence of instantaneous CSI. Instead, with stochastic CSI, given the large SNR, as discussed around Fig. \ref{Fig:CAPandCBP_C6T15NR4}, CBP is to be preferred. 

\begin{figure}[t]
\centering
\includegraphics[width=13cm]{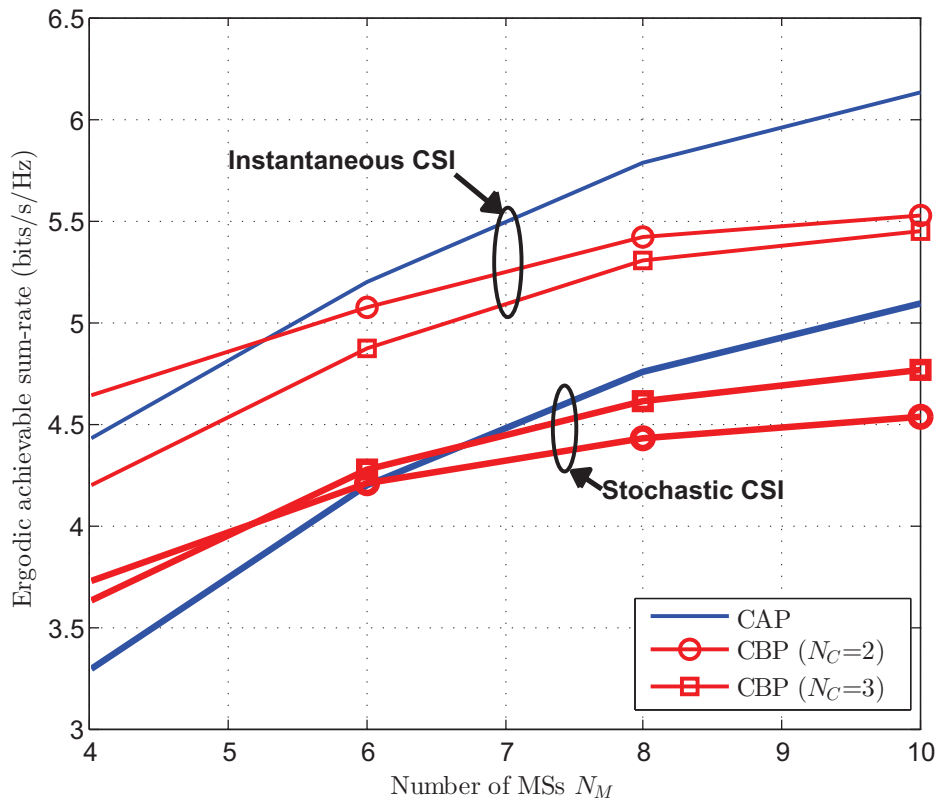}
\caption{Ergodic achievable sum-rate vs. the number of MSs $N_M$ ($N_R = 4$, $N_{t,i} = 2$, $N_{r,j} = 1$, $\bar C=4$ bits/s/Hz, $\bar P=10$ dB, $T = 10$, and $\mu=1$).}
\label{Fig:CAPandCBP_C4P10T10_NM}
\end{figure}

\begin{figure}[t]
\centering
\includegraphics[width=13cm]{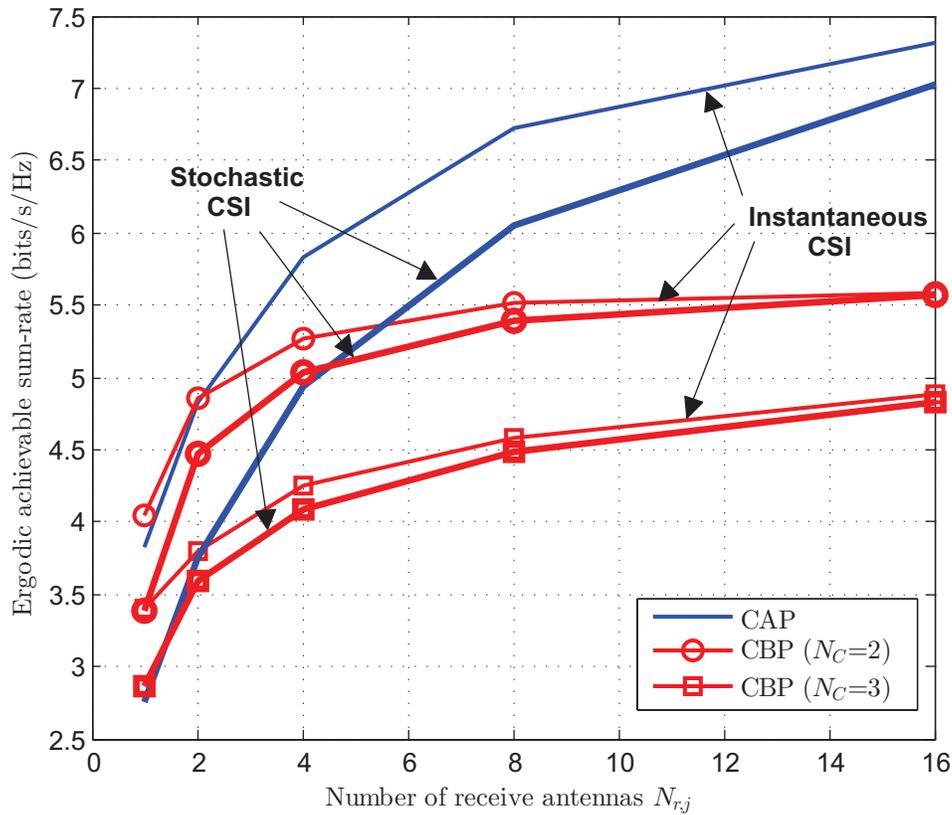}
\caption{Ergodic achievable sum-rate vs. the number of receive antennas $N_r$ ($N_R = N_M = 4$, $N_{t,i} = 2$, $\bar C=3$ bits/s/Hz, $\bar P=10$ dB, $T = 10$, and $\mu=1$).}
\label{Fig:CAPandCBP_C3P0T10_Nr}
\end{figure}



In Fig. \ref{Fig:CAPandCBP_C4P10T10_NM}, the ergodic achievable sum-rate is plotted versus the number of MSs $N_M$ for $N_R=4$, $N_{t,i} = 2$, $N_{r,j}=1$, $\bar C=4$, $\bar P=10dB$ and $T=10$. It is observed that the enhanced interference mitigation capabilities of CAP without the overhead associated to the transmission of all messages on the fronthaul links yield performance gains for denser C-RANs, i.e., for larger values of $N_M$. This remains true for both instantaneous and stochastic CSI cases.

Finally, in Fig. \ref{Fig:CAPandCBP_C3P0T10_Nr}, the ergodic achievable sum-rate is plotted versus the number of each receive antennas $N_{r,j}$ for $N_R = N_M = 4$, $N_{t,i} = 2$, $\bar C=3$ bits/s/Hz, $\bar P=10$ dB and $T=10$. Although the achievable rate of each MS is increased by using a large number of MS antennas, the achievable sum-rate with the CBP approach is restricted due to the limited number of cooperative RUs as dictated by the fronthaul capacity requirements for the transmission of the data streams. Hence, it is shown that the CAP approach provides significant advantages in the presence of a large number of antennas at MS for both instantaneous and stochastic CSI. Moreover, we observe that the performance advantages of having instantaneous CSI as compared to stochastic CSI decrease in the regime of the large number of MS antenna. This is because, in this regime, serving only one MS entails only a minor loss in capacity, hence not requiring sophisticated precoding operations. 
\section{Conclusion} \label{Sec:Conclusion}
In this paper, we have investigated the joint design of fronthaul compression and precoding for the downlink of C-RANs in the practically relevant scenario of block-ergodic fading with both instantaneous and stochastic CSI. The study compares the Compress-After-Precoding (CAP) and the Compress-Before-Precoding (CBP) approaches, which differ in their fronthaul compression requirements and interference mitigation capabilities. Efficient algorithms have been proposed for the maximization of the ergodic achievable sum-rate based on the stochastic successive upper-bound minimization technique. Extensive numerical results have quantified the regimes, in terms of fronthaul capacity, transmit power, channel coherence time and density of C-RANs, in which CAP and CBP are to be preferred. 

As a general conclusion, the relative merits of the two techniques depend on the interplay between the enhanced interference management abilities of CAP, particularly for dense networks, and the lower fronthaul requirements of CBP in terms of precoding information overhead, especially for large coherence periods and with stochastic, rather than instantaneous CSI. To elaborate, CBP requires data streams and precoding information to be sent on the fronthaul links. Hence, the fronthaul overhead of CBP increases with the network density, due to the larger number of data streams, and decreases with the coherence period and in the presence of stochastic CSI, owing to the reduced overhead for precoding. In contrast, the fronthaul overhead of CAP, which is due to the quantization of the baseband signals, does not depend on the network density, thus enabling to reap the interference management benefits of joint baseband processing at a larger scale. However, for small fronthaul capacities, large coherence periods and insufficiently dense networks, particularly in the presence of stochastic CSI, the interference management benefits of CAP may be outweighted by the lower fronthaul overhead of CBP. 
\newpage
\bibliographystyle{IEEEtran}
\bibliography{refKJK}

\end{document}